\documentclass[twocolumn,showpacs]{revtex4}[12pt]
\usepackage{graphicx}
\usepackage{color}
\begin{document}
\title{New electro-acoustic waves in a degenerate quantum plasma system}
\author{A. A. Mamun}
 \affiliation{Department of Physics,
Jahangirnagar University, Savar, Dhaka-1342, Bangladesh}
\date{\today}
\begin{abstract}
The existence of new electro-acoustic (EA) waves [named here `degnerate pressure driven EA'  (DPDEA) waves]  propagating  in a degenerate quantum plasma (DQP) system [containing  non-inertial, cold, 
non-relativistically (NR)/ultra-relativistically (UR) degenerate electron species (DES), and  inertial, cold, NR degenerate positive particle  species (PPS)]  is predicted for the first time.  The DPDEA waves, in which the inertia is mainly 
provided by the mass density of the inertial PPS, and  the restoring force is mainly  provided by the non-inertial, NR/UR DES, are new since  they are completely disappeared if the degenerate pressure of the plasma  
particle species is neglected.  The dispersion relation (derived here for the first time)  is applied in a white dwarf  DQP system to show the dispersion properties of the new DPDEA waves. 
\end{abstract}
\maketitle
The degenerate quantum plasma (DQP) systems  are  significantly  different from other plasma systems because of their extra-ordinarily high density and low temperature. These DQP systems having different unique properties 
do not only exist in space environments  \cite{Horn1991,Shukla2011a}, but also in laboratory devices \cite{Wolf2001,Atwater2007}.  They mainly contain non-inertial, non-relativistically/relativistically degenerate electron or/and positron \cite{El-Taibany2012}  species \cite{Horn1991,Shukla2011a,Wolf2001,Atwater2007}, and inertial, non-relativistically, non-degenerate/degenerate  ${\rm ~^{1}_{1}H}$  or ${\rm ~^{4}_{2}He}$ \cite{Horn1991} or ${\rm ~^{12}_{~6}C}$ or  
${\rm ~^{16}_{~8}O}$ \cite{Koester1990} or/and other heavy particle species \cite{Vanderburg2015,Witze2014}.
The degeneracy of the DQP particle species arises due to Heisenberg's uncertainty principle, and
that the  uncertainty in  momenta of extremely dense DQP plasma particle species  are infinitely large. This introduces a pressure called `degenerate pressure' which depends only  on the DQP particle number density, but not on thermal temperature. The degenerate pressure ${\cal P}$  exerted by any DQP particle species,  and  the corresponding degenerate energy ${\cal E}$ are given by \cite{Shukla2011b,MAS2016} 
\begin{eqnarray}
{\cal P}=Kn^{\gamma}, ~~~ 
{\cal E}=Kn^{\gamma-1},
\label{a}
\end{eqnarray}
where $n$ is the number density of any DQP particle species,  and  $\gamma$ and $K$ are defined as
\begin{eqnarray}
\gamma=\frac{5}{3}, ~~~
K=\frac{3}{5}(\frac{\pi}{3})^{\frac{1}{3}}\frac{\pi\hbar^2}{m}\simeq\frac{3}{5}\Lambda_{c}\hbar c,
\label{a1}
\end{eqnarray}
for non-relativistic limit  \cite{Shukla2011b,MAS2016}, and
\begin{eqnarray}
\gamma=\frac{4}{3}, ~~~
K=\frac{3}{4}(\frac{\pi^2}{9})^{\frac{1}{3}}\hbar c
\simeq\frac{3}{4}\hbar c,
\label{a2}
\end{eqnarray}
for ultra-relativistic  limit \cite{Shukla2011b,MAS2016}, where
$\Lambda_c=\pi\hbar/mc$, $\hbar$ is the reduced Planck constant,  and $m$ is the rest mass of any DQP particle species. We note that the subscript on the species-dependent  parameters and variables will be used $e$ or $j$ depending on the DQP particle species  $e$ or $j$.  This degenerate pressure ${\cal P}$ is balanced by either  self-gravitational pressure or electrostatic pressure depending on the DQP system under consideration to bring the DQP system into an equilibrium state.

The DQP systems are significantly different first because  they have unique properties (extra-ordinarily low density and temperature \cite{Brodin2017}), and next because they may  introduce new kind of waves. This  work is attempted to find the possibility for the existence of  a new kind of waves [called here `degenerate pressure-driven electro-acoustic (DPDEA) waves'  in which the inertia is mainly provided by the mas density of the inertial positive particle (PPS) species, and  the restoring force is mainly  provided by the non-inertial, ultra-relativistically (UR)/non-relativistically (NR) degenerate electron species (DES)], and to identify their dispersion properties in such DQP systems. 

We consider a DQP system containing non-inertial DES \cite{Horn1991,Wolf2001,Atwater2007} represented by the subscript $e$, and inertial, degenerate PPS represented by the subscript $j$,  which can be
${\rm ~^{1}_{1}H}$  or ${\rm ~^{4}_{2}He}$ \cite{Horn1991,Shukla2011a} or/and ${\rm ~^{12}_{~6}C}$ or  ${\rm ~^{16}_{~8}O}$ \cite{Koester1990} or/and other heavy particles \cite{Wolf2001,Atwater2007,Koester1990,Vanderburg2015,Witze2014}.  Thus, at equilibrium we have  $n_{e0}=Z_jn_{j0}$, where  $n_{e0}$ ($n_{j0}$) is the number density of the DES (PPS)  $e$  ($j$), and $Z_j$ is the charge state of the PPS $j$.  The dynamics of the new DPDEA waves, whose phase speed is much smaller than $C_{e}$ [where $C_{e}=({\cal E}_{e0} /m_e)^{1/2}$,  ${\cal E}_{e0}$  is the electron degenerate energy at equilibrium, and $m_e$ is the rest mass of an electron]  is described by
\begin{eqnarray}
&&\frac{\partial n_j}{\partial t} +\frac{\partial}{\partial x}(n_ju_j) = 0,
\label{be1}\\
&&\frac{\partial u_j}{\partial t} +u_j\frac{\partial u_j}{\partial x}=-\frac{Z_je}{m_j}\frac{\partial
\phi}{\partial x}+\frac{K_j}{m_j}\frac{\partial }{\partial x}(n_j^{\gamma_j-1}),
\label{be2}\\
&&\frac{{\partial}^2 \phi}{\partial x^2}=4\pi e \left[n_{e0}\left(1-\frac{e\phi}{{\cal E}_{e0}}\right)^{\frac{1}{\gamma_e-1}}-Z_jn_j\right],
\label{be3}
\end{eqnarray}
where $n_j$ ($u_j$) is the number density (fluid sped) of the inertial, degenerate PPS  $j$;  $\phi$ is the electrostatic wave potential; $m_j$ is the mass of  the inertial PPS  $j$;  $x$  ($t$) is the space  (time) variable, $e$ is the charge of a proton.  It may be noted here that the effect of the self-gravitational field is
neglected since  it is inherently  small in comparison with the electric field in the DQP system  (e. g.  white dwarf DQP system) under consideration. It should be mentioned here that the first term inside the square bracket  of 
(\ref{be3}) represents the expression for the number density of the DES, which is obtained by equating the electrostatic pressure to the degenerate pressure of the DES. 

To find the linear dispersion relation for these new DPDEA waves,  (\ref{be1})$-$(\ref{be3}) are linearized to 
reduce them to a set of linear equations:
\begin{eqnarray}
&&\frac{\partial \tilde{n}_j}{\partial t} +\frac{\partial}{\partial x} (n_{j0}\tilde{u}_j)= 0,
\label{L1}\\
&&\frac{\partial \tilde{u}_j}{\partial t} =-\frac{Z_je}{m_j}\frac{\partial
\tilde{\phi}}{\partial x}+ (\gamma_j-1)\frac{C_j^2}{n_{j0}}\frac{\partial
\tilde{n}_j}{\partial x} ,
\label{L2}\\
&&\frac{{\partial}^2 \tilde{\phi}}{\partial x^2}=4\pi e \left[\left(\frac{n_{e0}}{\gamma_e-1}\right)\frac{e\tilde{\phi}}{{\cal E}_{e0}}-Z_j\tilde{n}_j\right],
\label{L3}
\end{eqnarray}\
where $C_j= ({\cal E}_{j0}/m_j)^{1/2}$, ${\cal E}_{j0}$ is the equilibrium degenerate energy of  the degenerate PPS $j$;  
$\tilde{n}_j$,  $\tilde{u}_j$, and $\tilde{\phi}$ are the perturbed parts of the quantities involved. 

It is now assumed that
all of these perturbed quantities are directly proportional to
$\exp(-i\omega {t}+ikx)$, where $\omega$  ($k$) is the  angular
frequency (propagation constant) of the DPDEA waves. This
assumption reduces  (\ref{L1})$-$(\ref{L3}) to
 \begin{eqnarray}
\omega^2=\frac{(\gamma_e-1)k^2C_q^2+(\gamma_j-1)k^2C_j^2}{1+(\gamma_e-1)k^2L_q^2},
\label{dis}
\end{eqnarray}
where $C_q=(Z_j{\cal E}_{e0}/m_j)^{1/2}$ is the DPDEA wave speed, and $L_q=({\cal E}_{e0}/4\pi  n_{e0}e^2)^{1/2}$  is the DQP screening length.  It may be noted here that $\omega_{pj}=C_q/L_q$.  To examine the effect of degeneracy in PPS $j$ analytically,  (\ref{dis}) can be expressed as
\begin{eqnarray}
\omega=\frac{kC_q\sqrt{\gamma_e-1}}{\sqrt{1+(\gamma_e-1)k^2L_q^2}}(1+\beta_j)^{1/2},
\label{dis1}
\end{eqnarray}
where $\beta_j=(\gamma_j-1){\cal E}_{j0}/(\gamma_e-1)Z_j{\cal E}_{e0}$ represents the effect of the degeneracy in PPS $j$.  
To visualize the basic characteristics  of the new DPDEA waves, the values of ${\cal E}_{e0}$, $C_e$, $C_q$, $L_q$, ${\cal E}_{j0}/{\cal E}_{e0}$, $C_j/C_q$,  and $\beta_j$ for the DES  density 
as $n_{e0}=10^{29}~{\rm cm^{-3}}$, and for the PPS as ${\rm ~^{12}_{~6}C}$ are tabulated  in Table \ref{T1}.  
\begin{table}[htp]
\caption{Typical values of the physical quantities,
${\cal E}_{e0}$, $C_e$, $C_q$, $L_q$, ${\cal E}_{j0}/{\cal E}_{e0}$, $C_j/C_q$, and  $\beta_j$ for the DES  density as $n_{e0}=10^{29}~{\rm cm^{-3}}$, and for the PPS as ${\rm ~^{12}_{~6}C}$.}
\begin{tabular}{|c|c|c|c|}
\hline
Physical quantity                                            &NR DES                                     &UR DES \\
\hline 
${\cal E}_{e0}$ (ergs)                                    &${\rm 8.24\times 10^{-8}}$       &${\rm 1.46\times 10^-7}$ \\
\hline
$C_e$ (${\rm cm~s^{-1}}$)                           &${\rm 9.50\times 10^9}$           &${\rm 1.27\times 10^{10}}$\\
\hline
$C_q$ (${\rm cm~s^{-1}}$)                           &${\rm 1.56\times 10^8}$           &${\rm 2.10\times 10^8}$\\
\hline
$L_q~{\rm (cm)}$                                           &${\rm 5.32\times 10^{-10}}$     &${\rm 7.10\times 10^{-10}}$\\
\hline
${\cal E}_{j0}/{\cal E}_{e0}$                           &${\rm 1.37\times 10^{-5}}$       &${\rm 7.71\times 10^{-6}}$\\
\hline
$C_j/C_q$                                                        &${\rm 1.51\times 10^{-3}}$         &${\rm 1.13\times 10^{-3}}$\\
\hline
$\beta_j$                                                          &${\rm 2.30\times 10^{-6}}$       &${\rm 1.28\times 10^{-6}}$\\
\hline
\end{tabular}
\label{T1}
\end{table}
The limit $\beta_j\ll 1$, which  is a good approximation as obvious from Table I, leads  (\ref{dis1})  to 
\begin{eqnarray}
\omega=\frac{kC_q\sqrt{\gamma_e-1}}{\sqrt{1+(\gamma_e-1)k^2L_q^2}},
\label{dis2}
\end{eqnarray}
This is the dispersion relation for the DPDEA waves propagating in a DQP system (containing NR/UR DES, and non-degenerate inertial PPS), and   it is valid for both short and long wavelength limits.  The short-wavelength (compared to $L_q$)  limit ($kL_q\gg 1$) reduces  (\ref{dis1}) to $\omega=\omega_{jp}$, which is the upper limit of $\omega$, and is independent of the ultra-relativistic effect of the DES. On the other hand, the long  wavelength (compared to $L_q$) limit ($kL_q\ll 1$)  reduces  (\ref{dis1}) to $\omega=(\gamma_e-1)^{1/2}kC_q$. The latter can be expressed for NR and UR DES as
\begin{eqnarray}
\frac{\omega_{nr}}{k}=\sqrt{\frac{2}{3}}\sqrt{\frac{P_{e0}^{nr}}{\rho_{j0}}}=\sqrt{\frac{2\Lambda_{ce}\hbar c n_{e0}^{5/3}}{5n_{j0}m_j}}
 \end{eqnarray}
for NR DES,  and
\begin{eqnarray}
\frac{\omega_{ur}}{k}=\sqrt{\frac{1}{3}}\sqrt{\frac{P_{e0}^{ur}}{\rho_{j0}}}=\sqrt{\frac{\hbar c n_{e0}^{4/3}}{4n_{j0}m_j}}
 \end{eqnarray}
for UR DES, where $P_{e0}^{nr}$ ($P_{e0}^{ur}$) is the degenerate pressure exerted by NR (UR) DES,  
$\rho_{j0}=m_jn_{j0}$ is the mass density of the PPS, and $\omega_{nr}$ ($\omega_{ur}$) is the angular frequency of the DPDEA waves for NR (UR) DES. The last two equations mean that i) in  new DPDEA waves for the situation of NR (UR) DES,  the inertia is provided by the mas density ($\rho_{j0}$) of the inertial PPS, and  the restoring force is mainly  provided by the degenerate pressure  $ P_{e0}^{nr}$ ($P_{e0}^{ur}$)  exerted by the NR (UR)  DES, and ii) $\omega_{ur}/\omega_{nr}\simeq [5/(8\Lambda_{ce}n_{e0}^{1/3})]^{1/2}\simeq 1.05\simeq 1$. This means that for $kL_q\ll 1$, the UR DES does not have any effect on the  new DPDEA waves.

\begin{figure}[!h]
\centering
\includegraphics[width=8cm]{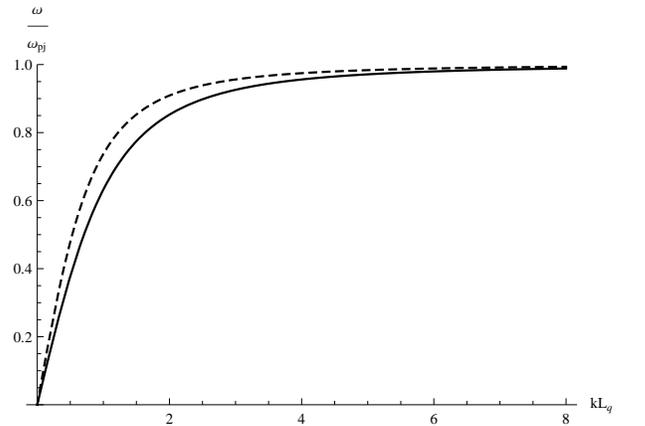}
\caption{The  dispersion curves of the DPDEA waves  for NR (solid curve) and  UR (dashed curve) DES, where  $L_q$ is for the NR DES.}
 \label{f1}
\end{figure}

We discussed the DPDEA waves for two extreme limits, viz.   $kL_q\ll 1$ and $kL_q\gg 1$, in which the UR DES does not have any effect on them. Now to examine the dispersion properties of them in between these two  limits, the dispersion relation (\ref{dis2}) have been numerically solved. The numerical results are displayed in figure \ref{f1},
which indicates that  the nature of the dispersion curve for the new DPDEA waves is similar to that for the well known ion-acoustic (IA) waves,  and  that the UR DES has insignificant effect on this new kind of DPDEA waves  for both short (compared to $L_q$) and long wavelength limits, but it has significant effect in between these two limits.  The UR DES enhances the phase speed of the DPDEA waves in between these two limits. It should be noted here that  the analytical analysis is completely agree with the numerical analysis displayed in figure \ref{f1} for these two limits. How the DPDEA waves are new, and different from the  IA  waves can be pinpointed in Table \ref{T2},
\begin{table}[htp]
\caption{The basic differences between the new DPDEA waves and the well-known IA waves.}
\begin{tabular}{|c|c|c|c|}
\hline
Properties                                                      &DPDEA Waves                                                                &IA Waves \\
\hline 
Restoring force                                              &Degenerate pressure                                                    &Thermal pressure \\
\hline
Phase speed                                                  &$\sqrt{\frac{{\cal E}_{e0}}{m_j}}$                                &$\sqrt{\frac{k_BT_e}{m_j}}$\\
\hline
Existence at $T_e=0$                                   &Possible                                                                         &Not possible\\                                                               
\hline
Length scale  (cm)                                         &$\sqrt{\frac{{\cal E}_{e0}}{4\pi  n_{e0}e^2}}$           &$\sqrt{\frac{k_BT_e}{4\pi  n_{e0}e^2}}$\\
\hline
\end{tabular}
\label{T2}
\end{table}
where $T_e$ is the electron temperature, and $k_B$ is the Boltzmann constant. We note that ${\cal E}_{e0}$, which is the equilibrium electron degenerate energy defined by (\ref{a})-(\ref{a2}), is completely independent of the thermal energy of any plasma species.  It is obvious from Table \ref{T2} that the DPDEA waves are new not only from the view of restoring force, but also from the view of their phase speed and length scale.

To conclude, the new results which have been found from this investigation can be pinpointed as follows:
\begin{enumerate}
 \item{The existence of a new kind of electro-acoustic waves (named here DPDEA waves) propagating in an degenerate plasma system is predicted.}
\item{The DPDEA waves are  new not only from the view of the restoring force (which is essential for the existence of any kind of waves),
but also from the view of their phase speed and length scale.} 
\item{The DPDEA  waves completely disappear if the degenerate pressure associated with any plasma species  is neglected.}
\item{The degenerate pressure associated with the inertial  PPS 
(providing here the inertia in the DPDEA waves) is negligible compared to that associated with the electron species (providing here the restoring force in the DPDEA waves).} 
\item{Unsatisfying  the condition $kL_q\ll1\ll kL_q$, the phase speed of the DPDEA waves  for the situation of UR DES is more than that in the situation of NR DES. This is because that the energy of the UR DES is more than that of the NR DES.} 
\end{enumerate}
The physics of the  DPDEA waves  is that if any column of any DQP medium is perturbed by any reason, i.e. the column of the medium is compressed (expanded), the degenerate  (electrostatic) pressure tries to bring it back to its equilibrium shape, but during this action, because of Newton's 1st law of motion and of inertial property of the PPS, it is expanded (compressed) more than its equilibrium shape,  it  is then compressed (expanded)   by the electrostatic (degenerate) pressure,  but  again during this action,  because of Newton's 1st law of motion and of inertial property of the PPS, it is  expanded (compressed) more than its equilibrium shape. This process will continue for infinite time to to produce and sustain the new DPDEA waves in the DQP system under consideration. Though the DPDEA wave dispersion relation (derived for the first time) is applied in white dwarf DQP system, it can be applied in any space and laboratory  DQP systems in understanding the basic features of the electrostatic perturbation mode in such  DQP systems.

\end{document}